# Highly Sensitive Differential Microwave Sensor Using Enhanced Spiral Resonators for Precision Permittivity Measurement

Nastouh Nikkhah, Rasool Keshavarz, Mehran Abolhasan, Justin Lipman, and Negin Shariati

*Abstract*— This paper presents a highly sensitive microwave sensor for dielectric sensing. One of the main disadvantages of microwave resonant-based sensors is cross-sensitivity originated by time-dependent uncontrolled environmental factors such as temperature that affect the material under test (MUT) behavior, leading to undesirable frequency shifts and, hence, lower accuracy. However, this work eliminates the unwanted errors using the differential measurement technique by comparing two transmission resonance frequencies during a unit test setup to measure the permittivity of MUT over time. The proposed structure comprises a spiral resonator with an extended horizontal microstrip line (EH-ML) coupled to a microstrip transmission line (MTL). Creating EH-ML within the structure comprises two primary contributions: enhanced sensitivity resulting from stronger fringing fields generated by increasing the effective area and improved resolution due to higher resonance frequencies caused by a lower total capacitive coupling effect. The proposed sensor is fabricated and tested using MUTs with a permittivity of less than 80 to verify the performance. In this regard, a frequency detection resolution (FDR) of 44MHz and a sensitivity of 0.85% are achieved at a maximum permittivity of 78.3. The results of theoretical analysis, simulation, and measurement are in relatively good agreement. Consequently, the proposed highly sensitive microwave sensor offers significant advantages, such as low complexity in design and fabrication. It also offers high resolution and precision in a wide range of permittivity, which can be an attractive candidate for dielectric sensing in health, chemical and agriculture applications.

*Index Terms*— Dielectric sensing applications, differential technique, microwave sensor, sensitivity, simple fabrication, spiral resonator.

## I. INTRODUCTION

MICROWAVE sensors (MSs) have been utilized for sensing through dielectric properties measurement of different materials in many applications such as blood glucose level (BGL) detection in biomedical [1]-[6], microfluidic systems [7]-[10], biosensors for health monitoring [11], agriculture [12]-[14] and chemistry [15], [16]. MSs have attracted significant interest among researchers due to significant advantages such as lightweight, low-cost, flexibility in design, simple fabrication, compactness, description of electromagnetic features of various materials [17], [18], real-time detection, easy integration to passive and active circuits [19], [20], [21] and application in the Internet of Things (IoT) devices [22], [23], [24]. Also, compatibility in the amount of coverage of material under test (MUT) compared to other equivalent models such as microelectromechanical (MEMs), thermal and optical devices, and the ability to achieve noticeable sensitivity and accuracy are other profits of MSs [25]. In addition, MSs can estimate the permittivity of the MUT merely by being in the vicinity, eliminating the need for direct contact. This contactless feature contributes to an extended sensor lifespan [26], [27].

Generally, the permittivity of MUT can be measured by MSs using coaxial, based on two main methods comprising broadband (non resonant-based) [28] and narrowband (resonant-based) [29] sensors. It should be noted that resonant-based models can lead to more accuracy than non resonant-based structures due to differences in the measurement process. The resonant-based microwave sensors evaluate the permittivity of MUT through changes in resonance frequency, phase or amplitude of S-parameters. On the other hand, the non resonant model measures permittivity according to impedance variations of MUT, which needs to be determined after any calibration, that can increase the probability of error and inaccuracy [30]. The permittivity of the MUT can be measured using common test setups such as vector network analyzer (VNA), software-defined radio (SDR), and sinusoidal or pulsed reference signals [31].

Microwave resonant-based sensors can be integrated with other sensor types to enhance the overall capabilities of the measurement system by providing a multifaceted approach to data collection [32]-[34]. One of the most important drawbacks of microwave resonant-based sensors is cross-sensitivity, which comprises measurement setup errors and unavoidable ambient

Manuscript received 20 November 2023. (Corresponding author: Nastouh Nikkhah). This work was supported by Food Agility Cooperative Research Centre (CRC) Ltd., through the Commonwealth Government CRC Program.

Nastouh Nikkhah is with the RF and Communication Technologies (RFCT) Research Laboratory, School of Electrical and Data Engineering, Faculty of Engineering and IT, University of Technology Sydney, Ultimo, NSW 2007, Australia, also with Food Agility CRC Ltd., Sydney, NSW 2000, Australia. (e-mail: Nastouh.nikkhah@student.uts.edu.au).

Rasool Keshavarz is with the RF and Communication Technologies (RFCT) Research Laboratory, School of Electrical and Data Engineering, Faculty of Engineering and IT, University of Technology Sydney, Ultimo, NSW 2007, Australia. (e-mail: Rasool.keshavarz@uts.edu.au).

Mehran Abolhasan is with the RF and Communication Technologies (RFCT)Research Laboratory, School of Electrical and Data Engineering, Faculty of Engineering and IT, University of Technology Sydney, Ultimo, NSW 2007, Australia. (e-mail: Mehran.Abolhasan@uts.edu.au).

Justin Lipman is with the RF and Communication Technologies (RFCT)Research Laboratory, School of Electrical and Data Engineering, Faculty of Engineering and IT, University of Technology Sydney, Ultimo, NSW 2007, Australia. (e-mail: Justin.Lipman@uts.edu.au).

Negin Shariati is with the RF and Communication Technologies (RFCT) Research Laboratory, School of Electrical and Data Engineering, Faculty of Engineering and IT, University of Technology Sydney, Ultimo, NSW 2007, Australia, and also with Food Agility CRC Ltd., Sydney, NSW 2000 Australia. (e-mail: Negin.Shariati@uts.edu.au).

circumstances. The permittivity of the MUT is a function of physical environmental parameters such as temperature and humidity, which are uncontrolled factors and may lead to unwanted frequency shifts and, hence, increased inaccuracy [35]. To reduce the unintentional errors caused by cross-sensitivity, some procedures, such as differential measurement techniques, can be reliable despite environmental factor changes that are effective on permittivity. Typically, differential-based microwave sensors need two sensing parts, leading to two or more resonance frequencies as they are exposed to similar environmental conditions [36],[37]. However, there are differential structures with only one sensing element based on the symmetry disruption concept, which is relatively insensitive to uncontrolled ambient factors [38]. Another disadvantage of resonant-based microwave sensors is a low Q-factor, leading to an insufficient resolution for sensing in some applications. In order to compensate for this issue, there are some strategies, such as using Metamaterial-inspired structures to obtain low-loss models resulting in a high Q-factor [12].

Many structures have evolved to provide planar microwave resonant-based sensors. The most popular planar resonators are first: 1) open-loop resonators (OLRs), 2) split ring resonators (SRRs) such as edge-coupled SRR (EC-SRR) or double-split SRR (2-SRR), and 3) complementary SRR (CSRR) [39]-[41]. It is worth mentioning that the OLRs and SRRs models can resonate independently [42] or coupled to a transmission line [43]. Also, the microwave resonators can be compact with high accuracy (high Q-factor) due to using the Metamaterial concept in their configurations [13]. Consequently, these contributions address a variety of sensing applications, such as biomedical, agricultural, and chemical systems. However, choosing an appropriate design is a function of operating frequency, required accuracy, type and volume of MUT, and permittivity range.

In this work, a lightweight, low profile modified 2-turn rectangular spiral resonators (M2TR-SR) sensor coupled at up ($U$) and down ($D$) of the microstrip transmission line (MTL) is designed, analyzed, simulated, fabricated, and measured. There is a good agreement among the proposed sensor's theory, simulation and measurement results. This structure includes three major elements: a conventional 2-turn rectangular spiral (C2TR-SR), extended horizontal microstrip line (EH-ML) and MTL, with a hollow bare FR4 container to keep MUT for the test. The proposed configuration involves a contactless microwave sensor which operates based on the near-field principle derived from the proximity of fringing electric fields and loaded MUT. The permittivity of MUT is measured through transmission resonance frequency shifts using a differential measurement technique. It should be noted that the printed circuit board (PCB) of the proposed sensor is coated with gold plating as a protective layer against corrosion, ultimately leading to an increase in the sensor's lifespan.

The main contributions of this work are summarized below:
- A conventional spiral resonator coupled to MTL resonates based on the mutual concept. Conducting an EH-ML leads to a higher sensitivity due to stronger fringing electric fields originated by a more effective area at the coupling section. Additionally, it enhances the resolution because of increased resonance frequency caused by a lower total capacitive coupling effect.
- The proposed sensor contains two sensing elements at up ($U$) and down ($D$) of the MTL in a unit structure which obtains two transmission resonance frequencies simultaneously. In this regard, unwanted frequency shifts created by time-dependent uncontrolled ambient factors are eliminated through differential sensing techniques during a unit measurement process of MUT permittivity at different times. Consequently, this work enhances accuracy.
- This work measures a wide range of the MUT dielectric constant (5 to 78.3) with a simple fabrication and low-cost test setup. It achieves a high sensitivity of 0.85% at the maximum permittivity, which represents the worst-case scenario, when compared to other findings listed in Table IV. Hence, the proposed sensor can be an appropriate candidate for dielectric sensing in many applications.

The rest of this paper is organized as follows. The theory, simulation analysis and application of this work are presented in Section II. Experimental results of the proposed sensor and comparison with other reported work are discussed in Section III. Finally, Section IV summarizes the major conclusions.

## II. THEORY AND SIMULATION ANALYSIS

This section is categorized into three parts. Initially, the principle of structure design using spiral resonators and coupling concepts will be discussed based on theoretical foundations and the proposed equivalent circuit model. Then, there is a comparison between the theory, circuit model and simulation results of the initial configuration. Further, the proposed sensor will be simulated and analyzed step-by-step following specific design elements. Finally, the application of this work in dielectric sensing will be analyzed.

### A. Sensor Design Theory

The schematic of a pair unit-cell M2TR-SR sensor comprising a C2TR-SR and EH-ML is illustrated in Fig.1 (a). There is an MTL that acts as a symmetrical load between a pair of the unit-cell M2TR-SR. The configuration, shown in Fig.1 (a), resonates at two independent frequencies through magnetic and electric edge-coupling the M2TR-SRs to the MTL. To clarify the concept of structure, the unit-cell equivalent circuit model is depicted in Fig.1 (b) [44] and [45]. When the width of MTL is wide, the mutual coupling between two M2TR-SRs can be ignored. Since the M2TR-SRs are small enough in terms of electrical size, the per-section inductance and capacitance of MTL can be modelled with the $L$ inductor and $C$ capacitor as lumped elements [46]. The $C_{SP\,(U,D)}$ and $L_{SP\,(U,D)}$ are the internal capacitance and inductance of the rectangular spiral. $C_{L\,(U,D)}$ and $L_{L\,(U,D)}$ are the capacitance created among EH-ML and MTL and the internal inductance of the EH-ML, respectively. $M_{(U,D)}$ is also the mutual inductance between M2TR-SR and MTL. According to [47], the equivalent circuit model can be simplified, as shown in Fig.1 (c),

$$C_i = \frac{L_{Si}}{M_i^2 \omega_i^2}, \qquad L_i = C_{Si} M_i^2 \omega_i^2, \qquad i = U, D \quad (1).$$

In this regard, the transmission response ($S_{21}$) of the two M2TR-SRs presents two resonance frequencies below,

$$\omega_i = \frac{1}{\sqrt{L_i C_i}} = \frac{1}{\sqrt{L_{Si} C_{Si}}}, \quad i = U, D \quad (2)$$

where $\omega_U$ and $\omega_D$ are resonance frequencies at up ($U$) and down ($D$) of the MTL, respectively. The proposed sensor is assumed to be lossless to simplify the description of the circuit and closed-form model. Furthermore, based on the sensor schematic in Fig.1 (a) and the equivalent circuit model in Fig.1 (b), which the EH-ML and C2TR-SR are in series, the $L_{Si}$, $C_{Si}$, $C_{Li}$, $L_{Li}$, $L_{SPi}$ and $C_{SPi}$ can be calculated below, respectively,

$$L_{Si} = L_{SPi} + \frac{L_{Li}}{2} \quad and \quad \frac{1}{C_{Si}} = \frac{1}{C_{SPi}} + \frac{1}{C_{Li}} \quad i = U, D \quad (3)$$

where,

$$C_{Li} = \left[\frac{A_{AVGi}}{G}\right] \varepsilon_{eff} \quad i = U, D \quad (4)$$

where the $A_{AVGi}$ and $G$ are the average effective area and gap distance among EH-ML and MTL, respectively. C2TR-SR is modified by creating the EH-ML ($C_{Li}$), which can enhance the sensitivity due to stronger fringing electric fields caused by a more effective area at the coupling part shown in Fig1. (a) (green arrow). Also, with emerging $C_{Li}$, the total capacitive coupling effect ($C_{Si}$) reduces, which leads to higher resonance frequency; consequently, the resolution of the structure is improved. In (5), Also, $\varepsilon_{eff}$ is the effective permittivity of microstrip structure with MUT on the top of it, which can be defined as [13],

$$\varepsilon_{eff} = \varepsilon_0 \left(\frac{(\varepsilon_S + \varepsilon_M)}{2} + \frac{(\varepsilon_S - \varepsilon_M)}{2} \frac{1}{\sqrt{1 + 12\frac{H_S}{W_{MS}}}}\right) \quad for \quad \frac{H_S}{H_M} \ll 1 \quad (5)$$

where, $\varepsilon_0$ is the dielectric permittivity coefficient of free space. Also, Fig.1 (d) shows that $\varepsilon_S$ and $\varepsilon_M$ are the low relative permittivity of the substrate and the dielectric constant of MUT which can be variable based on material type, respectively. In (6), the substrate thickness, MUT and the average width of the microstrip are expressed as $H_S$, $H_M$ and $W_{MS}$. It should be noted that the equations of $L_L$, $L_{SP}$ and $C_{SP}$ are discussed in the Appendix section.

All dimensions are presented based on international systems of units (SI). It is worth mentioning that all formulas mentioned above are general and can be applied for the sensor analysis in the next section.

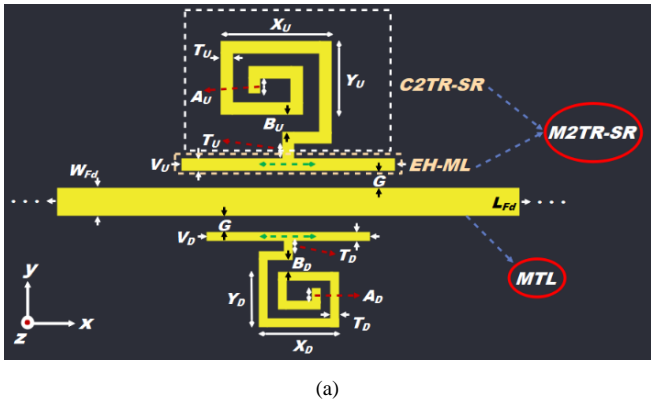

(a)

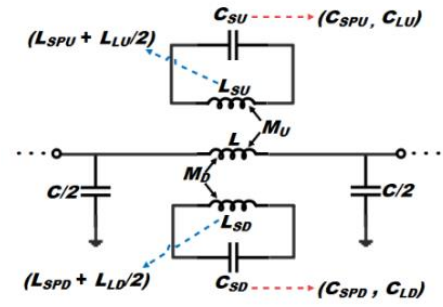

(b)

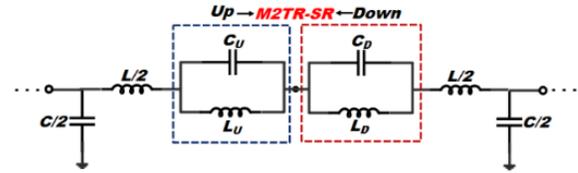

(c)

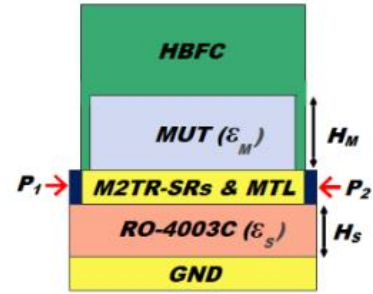

(d)

Fig.1. The proposed M2TR-SR sensor. (a) a pair of the proposed unit-cell schematic (top layer), (b) equivalent circuit model of the structure, (c) simplified equivalent circuit model using coupling concept, and (d) 3D layers of the system.

### B. Analysis of the Proposed Sensor

According to theoretical equations from (1) to (5) in Section II-part A and Table I, two transmission notch frequencies are obtained in a pair unit-cell M2TR-SR topology at up ($_U$) and down ($_D$) on MTL illustrated in Fig.1. It should be noted that theoretical equations are calculated using MATLAB software. First, in the bare (unloaded) case of MUT ($\varepsilon_B = 1$), the structure resonates at $f_{BU}$ =11.46 and $f_{BD}$ =17.55 GHz in up ($_U$) and down ($_D$) of the MTL. Also, two resonance frequencies of $f_{MU}$ =2.26 and $f_{MD}$ =3.17 GHz have been achieved when the dielectric constant or real part of permittivity ($\varepsilon'_M$) is 80. In order to support the theory, the transmission coefficient results ($S_{21}$), including electromagnetic wave and circuit model simulations, are illustrated in Fig.2. As can be seen in Fig.2 (a), two transmission zero frequencies are $f_{BU}$ =11.82 and $f_{BD}$ =17.89 GHz where MUT is air. On the other hand, two resonance frequencies of $f_{MU}$ =2.36 and $f_{MD}$ =3.2 GHz have been attained at $\varepsilon'_M = 80$, as shown in Fig.2 (b). Hence, a good agreement exists between full-wave simulation and equivalent circuit model results, which significantly support the theory. It should be noted that there is a slight difference in the $S_{21}$ amplitudes of the two plots in Fig.2 (a) since losses mainly created by substrate and MUT are not considered in the circuit model. Additionally, as shown in Fig.2 (a) on the left side, there is a smooth line between the two transmission notch frequencies at the unloaded condition ($\varepsilon_B =$ 1). However, as depicted in Fig.2 (a) on the right side ($\varepsilon'_M = 80$), some cluttered nulls between the transmission frequencies can arise when the MUT increases. This phenomenon can be attributed to the increased permittivity of the MUT, potentially leading to the emergence of unwanted resonator factors. Nevertheless, their impact on sensor performance is insignificant as long as the null depth of $|S_{21}|$ for the original resonant frequencies is adequate for detecting frequency shifts. The values of the equivalent circuit model parameters presented in Table II have been obtained through optimization using *CST Microwave Studio*.

TABLE I
DIMENSIONS OF THE PROPOSED SENSOR SHOWN IN FIG.1

| Unit: mm | X | Y | T | A | B | V |
|---|---|---|---|---|---|---|
| **UP** (U) | 1.7 | 1.3 | 0.2 | 0.3 | 0.2 | 3 |
| **Down** (D) | 1.19 | 0.91 | 0.14 | 0.21 | 0.14 | 2.1 |
| | $L_{Fd}$ | $W_{Fd}$ | G | $H_S$ | $H_M$ | |
| | 40 | 1.5 | 0.3 | 0.508 | 2 | |

At this stage, a few important design elements of the structure, including the number of cells and the distance between them, are analyzed. In this case, $|S_{21}|$ simulation results of the different number of cells ($N_C$ = 2 until 5) pairs of structures and centre-to-centre distance ($D_{CTC}$ = 10 to 18 mm) between them are illustrated in Fig.2 (b) and (c). The results confirm that the number and distance of cells would not be effective on the value of the resonance frequencies. However, according to [12] and Fig.2 (b) and (c), when the cell is added, the null depth of $|S_{21}|$ (sensor resolution concept which is equal to frequency shifts detection) and bandwidth of notch frequency increase. To check precisely, the 10-dB fractional bandwidth (FBW) and $|S_{21}|$ null depth variations based on the number of cells and $D_{CTC}$ at $\varepsilon'_M = 80$ are indicated in Fig.3 (a) and (b), respectively. It is important to highlight that in microwave sensors that measure the permittivity of MUT based on frequency shifts; the aim is to choose a lower FBW while achieving a greater null depth of $|S_{21}|$. As shown in Fig.3 (a) and (b), for 3-cell pair with a $D_{CTC}$ of 14 mm, FBW varies from 1 % to 1.5 %, and the $|S_{21}|$ null depth is approximately −20 dB. In conclusion, a tradeoff can be defined between the bandwidth and null depth of $S_{21}$. Due to the importance of dimensions in the design to achieve compact structure and based on simulation results in Fig.2 (b), (c) and Fig.3, the structure of a 3-cell pair with $D_{CTC}$ of 14 mm can lead to appropriate equilibrium among resolution and detection of $S_{21}$ variations.

TABLE II
AMOUNTS OF EQUIVALENT CIRCUIT MODEL SHOWN IN FIG.1

| MUT | C (pF) | L (nH) | $C_U$ (pF) | $L_U$ (pH) | $C_D$ (pF) | $L_D$ (pH) |
|---|---|---|---|---|---|---|
| $\varepsilon_B = 1$ | 0.02 | 0.2 | 19.77 | 9.15 | 14.75 | 5.37 |
| $\varepsilon'_M = 80$ | 0.2 | 0.1 | 45.05 | 101 | 47.55 | 52.3 |

Another design parameter is MUT thickness ($H_M$), which is optimally estimated at 2 mm in this work. However, the MUT thickness variations can slightly change the sensor behavior in frequency shifts and amplitude due to the effect of MUT on near-fields created by resonators and loss, respectively, which will be discussed in the next section. However, the MUT thickness changes would not affect the performance of the sensor. In this regard, Fig.4 depicts $|S_{21}|$ at $\varepsilon'_M = 10$ and $\varepsilon'_M = 80$ in the proposed sensor based on MUT thickness variations from 1.6 mm to 2.4 mm. Generally, electric field (E-Field) distribution is a bridge between theoretical concepts and practical observations, clearly showing how the presence of materials with different dielectric properties affects the sensor's E-Field distribution. In planar resonant-based microstrip structures, the resonance is characterized by a standing wave pattern in the E-Field distributions. The red color area in Fig.5 (a) at $\varepsilon_B = 1$ and Fig.5 (b) at $\varepsilon'_M = 80$ shows stronger electric fields (excitation through port 1) distributed on the resonators regarding the resonance frequencies. It should be noted the scale is presented logarithmically.

The structure is supplied using two discrete ports in *CST Microwave Studio,* and hence all simulation results in this section are considered on this basis.

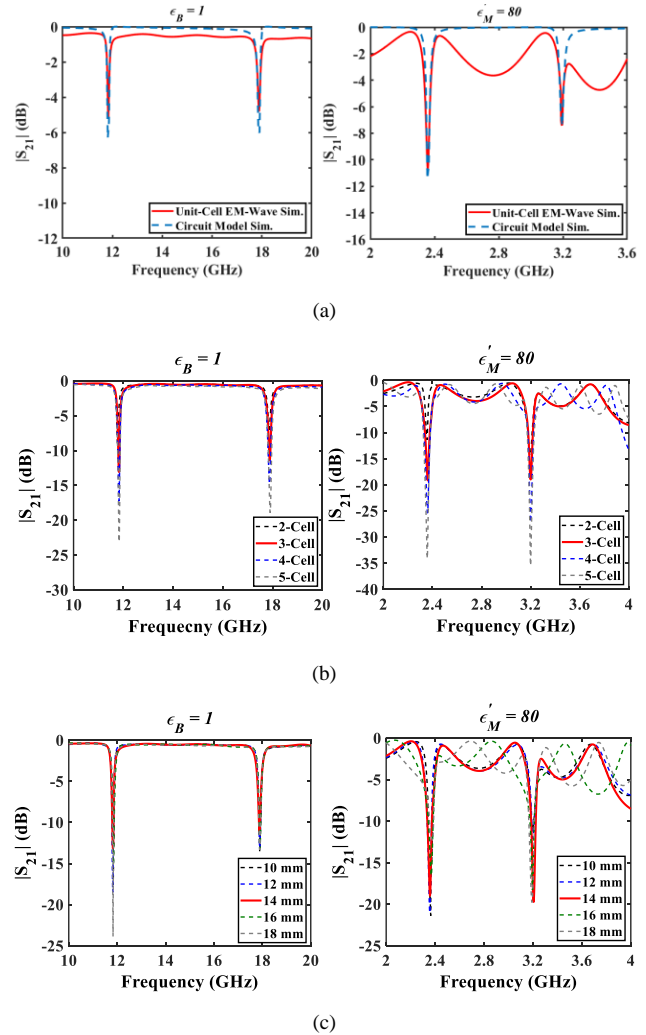

Fig.2. $|S_{21}|$ simulation results at unloaded ($\varepsilon_B = 1$) and worst-case scenario ($\varepsilon_M = 80$), (a) Comparison between unit-cell EM-Wave and proposed equivalent circuit model, (b) Number of cells from 2 to 5, (c) center to center distance between cells ($D_{CTC}$) from 10 mm to 18 mm.

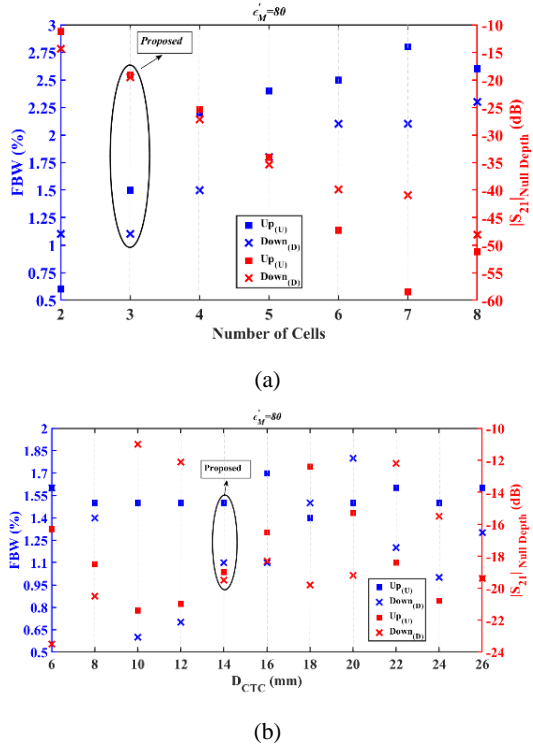

Fig.3. Simulated results of FBW (%) and $|S_{21}|$ (null depth) arising from cells at up ($U$) and down ($D$) of MTL in $\varepsilon_M = 80$ based on (a) the number of cells (2 to 8 cells), (b) center to center distance ($D_{CTC}$) between cells from 6 mm to 26 mm (3 cells).

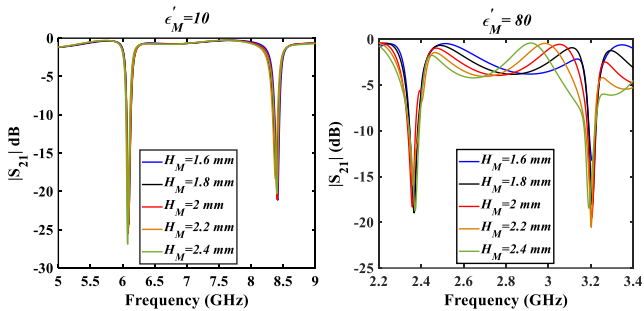

Fig.4. Simulated results of $|S_{21}|$ for different thicknesses ($H_M$) of MUT at two samples of $\varepsilon'_M=10$ and $\varepsilon'_M=80$.

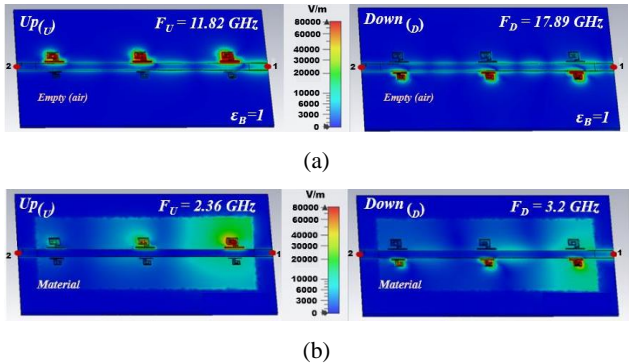

Fig.5. E-field distribution of the proposed sensor (a) at $\varepsilon_B = 1$, (b) $\varepsilon'_M = 80$.

### C. Application in Dielectric Sensing

In Section II-part B, the M2TR-SRs using MTL load results in two isolated notch frequencies in transmission response. This sensor has been proposed to sense different materials by detecting frequency shifts of $S_{21}$, which can be altered based on the loaded MUT dielectric constant changes. This concept can be analyzed using field and theory viewpoints. From field pinpoint, regarding interplay among fringing electric fields of M2TR-SRs and loaded MUT, two zero frequencies of transmission response shift. This hypothesis originates from the near-field theory, where the fringing electric fields of M2TR-SRs and loaded MUT are adjacent [51]. It should be mentioned that the resonance frequency of both M2TR-SRs at up and down of MTL can change. To address this, the size of the MUT part is 34mm × 15mm × 2mm to encompass the whole resonator area. Furthermore, theoretically, two notch frequencies can be modified through M2TR-SRs inductance ($L_{Li}$ and $L_{Spi}$) and capacitance ($C_{Li}$ and $C_{Spi}$) variations in (3). Also, as described in (6) and (9), the $C_{Li}$ and $C_{Spi}$ function the effective permittivity ($\varepsilon_{eff}$) which is dependent on the value of $\varepsilon_M$ changes based on (5). On the other hand, the $L_{Li}$ and $L_{Spi}$ alterations are not effective in the $S_{21}$ frequency shift; hence, they are constant parameters in the analysis and measurement process. According to equation (2), the notch resonance frequency can be derived below as a function of the permittivity of MUT ($\varepsilon_M$) whereas $\theta_i$ is considered a constant parameter,

$$f_i = \frac{1}{2\pi\sqrt{L_{Si}C_{Si}}} = \frac{1}{2\pi\sqrt{\theta_i \varepsilon_M}} \quad i = U, D \quad (6)$$

where according to equation (3),

$$L_{Si}C_{Si} = L_{Si}\frac{C_{Li} \times C_{SPi}}{C_{Li} + C_{SPi}} \to \theta_i = L_{Si}\frac{\beta_{1i} \times \beta_{2i}}{\beta_{1i} + \beta_{2i}}\alpha_{1i} \quad i = U, D \ (7)$$

which $L_{Si}$ is referred to by (3) and $\beta_{1i}$ and $\beta_{2i}$ are represented in $C_{Li} = [\beta_{1i}]\varepsilon_{eff}$ and $C_{SPi} = [\beta_{2i}]\varepsilon_{eff}$ by (4) and (A4) concurrently. Additionally, $\alpha_{1i}$ is defined as a constant coefficient in the form of $\varepsilon_{eff} = \alpha_{1i}\varepsilon_M$ derived by (5) and calculated by MATLAB software. The most important factor in MS applications is sensitivity. In this work, two zero frequencies of $S_{21}$ shift based on $\varepsilon_M$ changes. The frequency detection resolution (FDR) indicates the sensitivity only regarding load variations. However, as a more precise definition, the concept of sensitivity would be independent of unloaded frequency ($f_B$) due to the non-linear behavior of frequency changes over the $\varepsilon_M$ range. In this regard, according to [52], the FDR and sensitivity can be estimated below,

$$FDR_i = \frac{\partial f_{Mi}}{\partial \varepsilon_M} = \lim_{\varepsilon_B \to \varepsilon_M} \frac{f_{Bi} - f_{Mi}}{\varepsilon_B - \varepsilon_M} \quad i = U, D \quad (8)$$

where $f_{Bi}$ and $f_{Mi}$ are zero frequencies of $S_{21}$ at unloaded ($\varepsilon_B$) and loaded ($\varepsilon_M$) states, respectively. Hence, by derivation from equation (6),

$$FDR_i = \frac{1}{2\pi\varepsilon_M\sqrt{\theta_i \varepsilon_M}} \quad i = U, D \quad (9).$$

Equation (9) indicates that if the dielectric constant of MUT increases, the FDR and sensitivity will be decreased. In addition, based on [53], the normalized sensitivity is calculated as follows,

$$S_i(\%) = \frac{FDR_i}{f_{Bi}} \times 100 \quad i = U, D \quad (10).$$

Furthermore, to clarify the differential sensing process as an example, the unwanted frequency shift created by time-dependent uncontrolled environmental factors such as temperature and humidity is supposed $\kappa$ at two different times ($t_1$ and $t_2$), which is shown below,

$$\begin{cases} t_1: f_{1D}, f_{1U} \to DIFF = f_{1D} - f_{1U} \\ t_2: f_{1D} + \kappa, f_{1U} + \kappa \to DIFF = f_{1D} - f_{1U} \end{cases} \quad (11)$$

where $f_{1D}$ and $f_{1U}$ describe the transmission resonance frequencies at $t_1$. Based on (11), $DIFF$ (differential value) is constant and therefore, error ($\kappa$) is removed, which leads to improving accuracy. On the other hand, the proposed FDR ($FDR_P$) is defined below,

$$FDR_p = \frac{D_P}{\Delta \varepsilon}, \quad \Delta \varepsilon = \varepsilon_B - \varepsilon_M \quad (12)$$

which $D_P$ is inferred based on the differential measurement technique given by,

$$D_P = |(f_{MD} - f_{MU}) - (f_{BD} - f_{BU})| \quad (13)$$

also, the values obtained from (13) can be simply equal to $FDR_D - FDR_U$ and ultimately, the proposed sensitivity can be defined as follows,

$$S_P(\%) = \frac{FDR_p}{\Delta f_B} \times 100 \quad (14)$$

where, $\Delta f_B$ describes the difference between two resonance frequencies of $S_{21}$ at bare condition. Another factor in the performance of the sensor is the quality factor. The quality factor is given as $Q_{factor} = \frac{f_n}{f_{3dB}}$, where $f_n$ and $f_{3dB}$ are notch frequency and −3 dB bandwidth of $S_{21}$, respectively. To validate theory and simulation results through measurement, a hollow bare FR4 container depicted in Fig.7 (a) with the size of 35mm × 19mm × 2mm is designed to maintain materials for the test. Since this container would not be effective in the desired results, the $S_{21}$ simulation results of the proposed sensor with HBFC are indicated in Fig.6 (b) and (c) for validation. As can be seen in Fig. 6 (b) and (c), there is no difference without and with HBFC. Remarkably, this container can be an appropriate candidate for functional measurement setup due to the simplicity of implementation, variety in the choice of materials and usability for many applications. Moreover, as shown in Fig.6 (a), two SMA connector models are added to the simulation. It is inevitable that adding an SMA connector leads to increasing loss at high frequencies, which is related to the amplitude of $S_{21}$, and hence, two transmission resonance frequencies do not change. It should be noted that the loss of MUT is ignored in Fig.6 (b) and (c) since it is not significantly effective on resonance frequency shifts in the microstrip structures, according to [54].

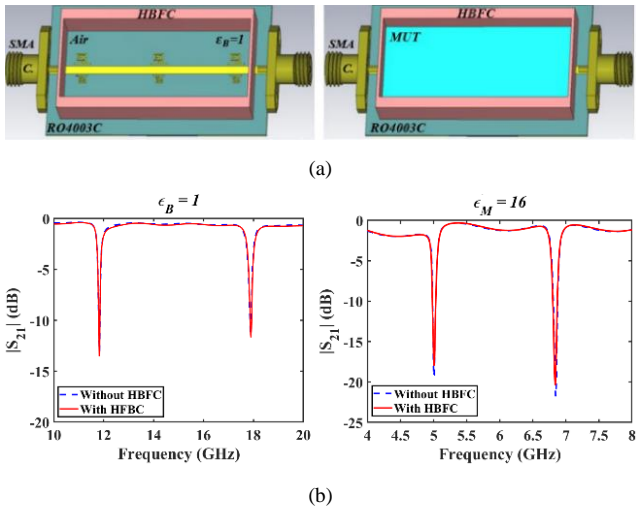

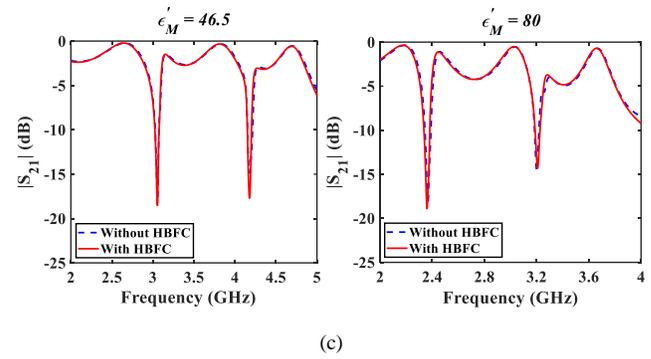

(c)

Fig.6. (a) 3D views of the proposed schematic structure with a container at empty and loaded conditions, simulated of $|S_{21}|$ with and without container at (b) $\varepsilon_B = 1$ & $\varepsilon'_M = 16$ (c) $\varepsilon'_M = 46.5$ & $\varepsilon'_M = 80$.

## III. EXPERIMENTAL RESULTS AND DISCUSSION

In order to validate the theory and simulation through measurement, a proposed M2TR-SR sensor is fabricated on Rogers RO4003C substrate with a thickness of 0.508 mm, a dielectric constant of 3.55 and a loss tangent of 0.0027 shown in Fig.7 (a). Two SMA connectors are connected to MTL through two input ports to feed the sensor. The gold plating is performed on the proposed sensor to prevent corrosion of copper during tests by different MUTs, which increases the sensor's lifespan. Also, a hollow bare FR4 container with a volume of 0.99 cc is fabricated, as indicated in Fig.7 (b), to keep the MUT. This container is installed on the sensor through a strong drop glue without any leakage towards SMA connectors and feed points, which should be completely dry since input matching impedances should not vary when using liquids such as ethanol or water as a MUT in measurement setup tests. Measurements are performed using VNA-ZVA40.

Fig.8 (a) and (b) illustrate the measurement results of two zero frequencies, which are 12.09 GHz ($_U$) and 17.22 ($_D$) GHz at $\varepsilon_B$=1 (air). Therefore, according to the differential method, the resonance frequency is considered $\Delta f_{B(Measurement)}$=5.13 GHz at bare (unloaded) condition. Further, the theory and simulation results are 11.46 GHz, 17.55 GHz in analysis ($\Delta f_{B(Theory)}$=6.09 GHz) and 11.82 GHz and 17.89 GHz in simulation ($\Delta f_{B(Simulation)}$=6.07 GHz) at up ($_U$) and down ($_D$) cells, respectively. The results obtained from theory, simulation, and measurement demonstrate a good agreement. It should be noted that the variations of MUT behavior at the unloaded condition (air) caused by uncontrolled environmental factors such as temperature and humidity are not tangible at different frequencies; therefore, the results of the differential process would be valid in the proposed sensor. In the measurement tests, the value of $f_{BD}$ is observed to be lower than the theoretical and simulated results, which can be attributed to the parasitic effects resulting from the length of the center male pin of the SMA connector, as well as the impact of soldering at high frequencies. As mentioned in Section II-part C, based on the fringe fields notion, resonance frequency variations can lead to dielectric sensing through the real part of permittivity measurement ($\varepsilon'_M$). Hence, to validate simulation results, soil with a permittivity of about 5 is considered as a MUT [12]. In this case, shown in Fig.8 (a) and (c), two resonance frequencies are obtained at $f_{MU}$ =8.65 GHz and $f_{MD}$ =11.25 GHz. Also, the $FDR_P$ and sensitivity are 0.627 GHz and 10.6% concurrently.

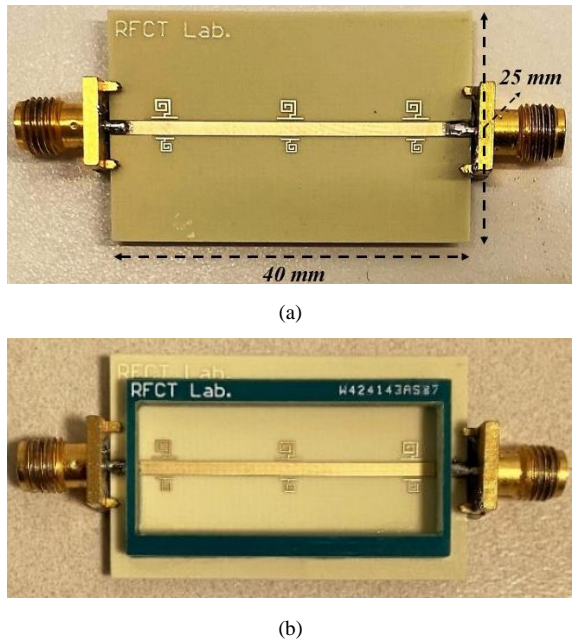

(a)

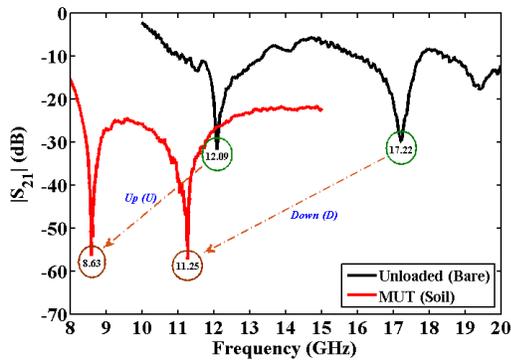

(b)

Fig.7. (a) Prototype of the proposed sensor implemented on Rogers RO4003C with $\varepsilon_r$ = 3.55, thickness 0.508 mm, and loss tangent 0.0027, (b) Installed the hollow bare FR4 container.

(a)

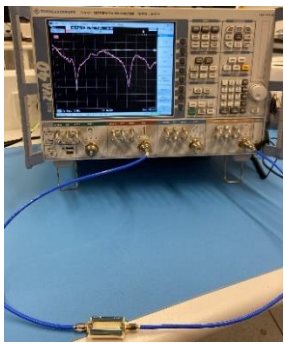 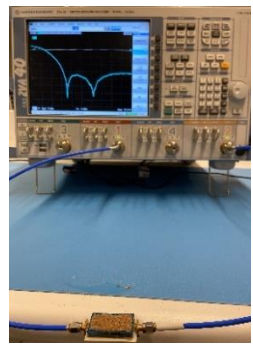

(b)          (c)

Fig.8. $|S_{21}|$ measurement results of the proposed sensor (a) real permittivity variations of MUT including bare and soil, leading to frequency shift, (b) measurement setup (MUT: air), (c) test setup (MUT: soil).

Furthermore, there are remarkable applications in health, agriculture, and chemical sectors at high dielectric constants, specifically until about 80. In this case, to analyze the sensor's functionality, the proposed sensor is also tested through different MUTs, as shown in Table III. All values are derived through references [12], [55]-[57] and verified by the Dielectric Assessment Kit (DAK-12) device [58].

TABLE III
AMOUNTS OF MUT DIELECTRIC CONSTANT TESTED ON SENSOR

| REF | MUT | Temperature | $\varepsilon'_M$ |
|---|---|---|---|
| [12] | Soil with 30% VWC | 25 °C | 16 |
| [55] | Ethanol | | 25.3 |
| [56] | Glycerol | | 46.5 |
| [57] | Water | 100 °C | 55.7 |
| | | 75 °C | 62.4 |
| | | 60 °C | 66.8 |
| | | 45 °C | 71.5 |
| | | 37 °C | 74.5 |
| | | 25 °C | 78.3 |

The measurement results of $S_{21}$ with test setup when MUT permittivity changes from 55.7 to 78.3 are indicated in Fig.9 (a) and (b). Transmission resonance frequencies are $f_{MU}$ =1.688 GHz and $f_{MD}$ =3.445 GHz at the maximum value of $\varepsilon'_M$ that is 78.3. One of the important points in measurement results is loss clearly affected by MUT at high frequencies in Fig.8 (a) and (c), which can lead to less amplitude (high depth null) such as −60 dB. This phenomenon is not simulated precisely due to software limitations in processing. However, achieving zero transmission frequencies plays a key role in the measurement process, while amplitude cannot be important since the proposed structure offers MUT permittivity sensing based on resonance frequency shifts.

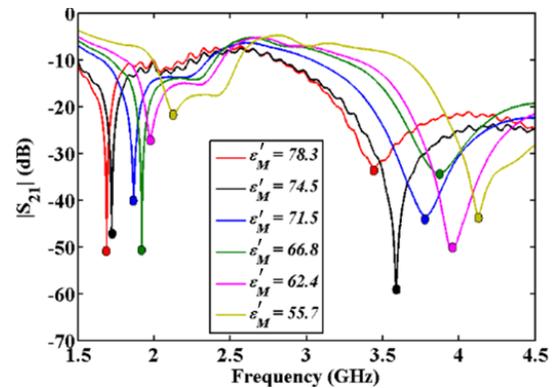

(a)

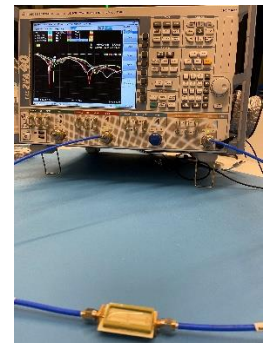

(b)

Fig.9. Measured results of $|S_{21}|$ (a) real permittivity variations of MUT from $\varepsilon'_M$=55.7 to $\varepsilon'_M$=78.3 leading to frequency shift, (b) test setup (MUT: water at different degrees).

In order to validate the sensor's performance, $FDR_P$ and proposed sensitivity are depicted in Fig. 10 (a) and (b) based on theory, simulation and measurement results in real permittivity within a range of 5 to 78.3. In this case, the test results, including 44 MHz of $FDR_P$ and 0.85% of proposed sensitivity, are attained in the worst-case scenario, which is $\varepsilon'_M = 78.3$. As can be seen in Fig. 10 (a), (14) and (15), when the dielectric constant of MUT ($\varepsilon'_M$) increases, the $FDR_P$ and sensitivity reduce. Also, the $FDR_P$ and $S_P$ results show a sudden decrease in less than approximately $\varepsilon'_M=25$, and then changes are relatively stable. Furthermore, in Fig.10 (a) and (b), mathematical models describe the measured values of $FDR$ ($y_{FDR}$) and proposed sensitivity ($y_{S_P}$) as a function of permittivity of MUT ($x$) using 1$^{th}$-term power nonlinear equations through curve fitting in MATLAB and given by,

$$y_{FDR} = ax^b \to a = 3.09 \ \& \ b = -0.9926 \ (R^2 = 0.9994) \quad (15)$$

$$y_{S_P} = ax^b \to a = 46.62 \ \& \ b = -0.92 \ (R^2 = 0.9997) \quad (16).$$

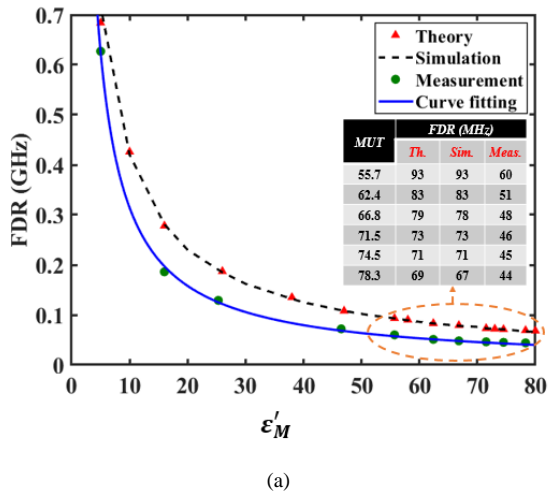

(a)

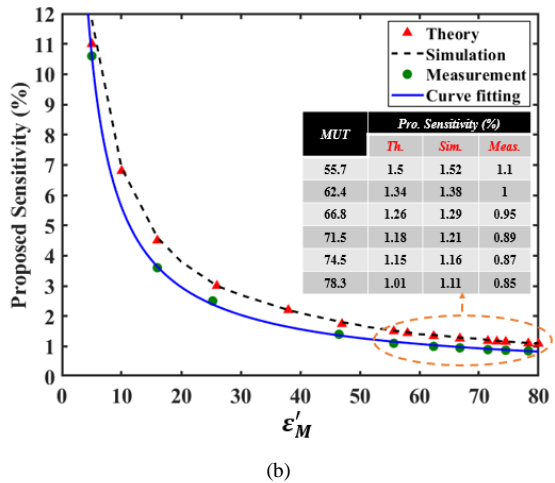

(b)

Fig.10. Comparison of theory, simulated, and measurement results with curve fitting in a wide range of MUT permittivity from 5 to 78.3, (a) $FDR_P$ calculated by (12), and (b) $S_P$ derived by (14).

To verify the performance of the sensor's ability to eliminate unwanted frequency shifts caused by time-dependent uncontrolled environmental factors through differential process, the changes of $S_{21}$ are measured using the Environmental Test Chamber (Vötschtechnik®). As a proof of concept, Fig. 11 shows the test setup while MUT is ethanol at 25°C and water at the two temperatures, 25°C and 37°C, for 3 minutes (start time: 0 / stop time: 3 minutes). The chamber is adjusted at 25°C and 37°C (± 2°C tolerance), respectively, with the same humidity of 50% (± 2% tolerance). The ±2°C and ±2% tolerances can create similar environmental conditions in the open-space tests. According to the test setup illustrated in Fig. 11, the uncontrolled variations of $f_{1D}$ and $f_{1U}$ are approximately 8 MHz for the ethanol state and 2 MHz at 25°C as well as 3 MHz at 37°C for the water state, which can be removed through the differential process; therefore, the $DIFF$ at (11) does not change resulting in higher accuracy has been achieved.

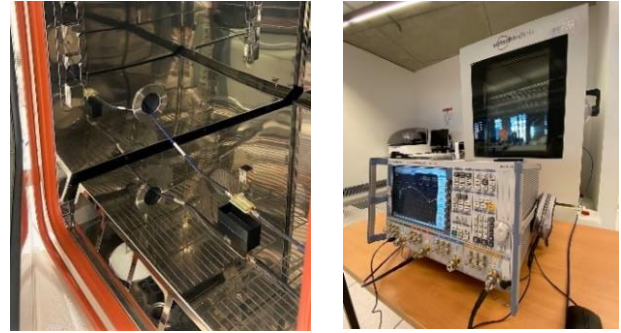

Fig.11. Test setup of sensor performance using Environmental Test Chamber (MUT: ethanol and water at two different $\varepsilon'_M$).

Finally, the proposed sensor is compared with other reported works in Table IV. Almost all state-of-art sensors are resonator-based and function using the frequency shift technique. Since the microwave sensor's behavior and challenges are different at a high permittivity than at a low one, all reported works have been compared at the highest permittivity, approximately 80. Two crucial characteristics can be addressed in the comparison table. First, the differential process as a remarkable method to reduce the undesirable environmental effects of the test setup, which are utilized in [9], [64] and the proposed sensor in this work. Further, sensitivity is a critical parameter in sensor functionality. As can be seen in Table IV, the proposed sensitivity in this work is 0.85% at maximum permittivity, which is higher than [59], [60], [62], [63], and [64], also more than twice compared with [9],[10],[25],[40], and [61]. Consequently, the proposed sensor can be an promising candidate for dielectric constant sensing applications such as non-invasive biomedical, agricultural and chemical systems.

## IV. CONCLUSION

A highly sensitive microwave resonant-based sensor with a differential measurement process for dielectric sensing applications using a modified spiral resonator coupled to a MTL is designed, analyzed, simulated and tested. The proposed structure with a dimension of 40 mm×25 mm determines the dielectric constant of MUT from 5 to 78.3 using a simple container based on transmission resonance frequency variations. Making an extended horizontal microstrip line to a spiral resonator enhances sensitivity due to a more effective area with MTL. This leads to stronger fringing fields and improves the resolution arising from higher resonance frequency caused by a lower total capacitive coupling. This sensor removes unwanted frequency shifts created by time-dependent unmanageable environmental factors such as temperature and humidity using the differential measurement process for the dielectric constant of MUT over time, leading to higher accuracy. Also, the structure is not sensitive to the MUT thickness in a specific range (1.6 to 2.4 mm). In conclusion, the proposed sensor addresses a low-cost, lightweight, simple fabrication, good accuracy with significant sensitivity (0.85 % at real permittivity of 78.3) compared to other reported publications. The proposed sensor can be used in a wide range of dielectric sensing applications.

## ABBREVIATION

A glossary of the mentioned terms is shown below:

| Acronym | Description |
|---|---|
| MS | Microwave sensors |
| MUT | Material under test |
| C2TR-SR | Conventional 2-turn rectangular spiral resonator |
| EH-ML | Extended horizontal microstrip line |
| MTL | Microstrip transmission line |
| M2TR-SR | Modified 2-turn rectangular spiral resonator |
| FBW | Fractional bandwidth |
| $S_{21}$ | Transmission coefficient from port 1 to port 2 |
| VNA | Vector network analyzer |
| FDR | Frequency detection resolution |
| $S_P$ | Proposed sensitivity |

## APPENDIX

In the sensor theory design section, the $L_{Li}$ is estimated [48],

$$L_{Li(\mu H)} = 0.2 V_i [ln\frac{2V_i}{H_S+T_i} + 0.5 + 0.22(\frac{H_S+T_i}{V_i})] \quad i = U, D \quad (A1),$$

Moreover, the $L_{SPi}$ is expressed [49],

$$L_{SPi} = \mu_0 \frac{N^2}{H_S} A_{SPi} \quad if \quad (A_{SPi} = X_i Y_i) \quad i = U, D \quad (A2),$$

It should be noted that $\mu_0$ is permeability in free space, and the number of turns of the spiral describes *N*. As mentioned above, the capacitor of a spiral structure is defined $C_{SPi}$ and given by [50],

$$C_{SPi} = \left[\frac{K_i(\sqrt{1-\gamma^2})}{K_i(A_i)} Q_i\right] \varepsilon_{eff} \quad for \quad \gamma = \frac{\frac{B_i}{2}}{\frac{B_i}{2}+T_i} \quad i = U, D \quad (A3),$$

Which $K_i$ is the complete elliptic integral of the first kind, and also,

$$Q_i = \frac{X_i}{4(G_i+T_i)} \frac{N^2}{N^2+1} \left[X_i - (N - \frac{1}{2})(B_i + T_i)\right] \quad i = U, D \quad (A4).$$

## ACKNOWLEDGMENT

This project was supported by Food Agility CRC Ltd, funded under the Commonwealth Government CRC Program. The CRC Program supports industry-led collaborations between industry, researchers and the community.

TABLE IV
COMPARISON OF THE PROPOSED SENSOR AND OTHER REPORTED WORKS

| Ref. | Sensing technique | Unloaded frequency (GHz) | Size $(\lambda_0)^2$ | Req. volume (mm³) | Q-factor at unloaded condition | FDR (MHz) at maximum $\varepsilon'_M$ | Sensitivity (%) at maximum $\varepsilon'_M$ | Design complexity | Differential |
|---|---|---|---|---|---|---|---|---|---|
| [9] | Planar TL + series RLC resonator | 1.98 | - | 0.46 | - | 5.9 | 0.3 | Moderate | Yes |
| [10] | Central gap ring resonator | 2.53 | 0.59×0.31 | 13.3 | 2993 | 1.3 | 0.05 | High | No |
| [25] | MTM-based | 2.6 | - | 5 | - | 7 | 0.27 | Moderate | No |
| [40] | CSRR | 2.16 | 0.42×0.14 | 400-1200 | - | 4.2 | 0.21 | Low | No |
| [59] | SRR | 1.72 | 0.37×0.17 | 0.68 | 97.6 | 14.1 | 0.82 | Moderate | No |
| [60] | Complementary Electric-LC and SRR | 2.45 | 0.32×0.2 | 1.67 | 201.8 | 15.9 | 0.65 | High | No |
| [61] | Square SIW re-entrant cavity resonator | 2.19 | 0.4×0.36 | 35.9 | 1700 | 8 | 0.366 | High | No |
| [62] | Meander open CSRR | 0.33 | 0.05×0.03 | 80 | - | 1.7 | 0.504 | Moderate | No |
| [63] | Series LC-based | 1.662 | 0.2×0.1 | 0.7 | - | 11.5 | 0.695 | High | No |
| [64] | MCSRR | 1.62 | 0.44×0.28 | - | - | 7.6 | 0.469 | Low | Yes |
| This work | M2TR-SR | 5.13 | 1.74×1.08 | 990 | 12 | 44 | 0.85 | Low | Yes |